\documentclass[journal]{IEEEtran}

\usepackage{epsfig,amsmath,amssymb,epsf,amsthm,scalefnt,multirow}
\usepackage{xcolor}
\usepackage{float}
\usepackage{cite}
\usepackage{subcaption}
\usepackage{setspace}
\newcommand{\argmax}{\mathop{\mathrm{argmax}}}
\newcommand{\argmin}{\mathop{\mathrm{argmin}}}

\newtheorem{theorem}{Theorem}
\newtheorem{lemma}{Lemma}

\newtheorem{corollary}{Corollary}

\newtheorem{proposition}{Proposition}

  \def\cC{{\mathcal{C}}} 
 \def\cF{{\mathcal{F}}}  
   
 \def\cN{{\mathcal{N}}}

\def\argmin{\mathop{\mathrm{argmin}}}
\def\argmax{\mathop{\mathrm{argmax}}}
\def\diag{\mathop{\mathrm{diag}}}
\def\inf{\mathop{\mathrm{in}}}
\def\outf{\mathop{\mathrm{out}}}

\def\Re{\mathop{\mathrm{Re}}}

\def\bSigma{{\pmb{\Sigma}}}

\def\b0{{\pmb{0}}} 

 \def\bb{{\mathbf{b}}} \def\bc{{\mathbf{c}}} 
 \def\bff{{\mathbf{f}}} \def\bg{{\mathbf{g}}} \def\bh{{\mathbf{h}}}
   
 \def\bn{{\mathbf{n}}}  \def\bp{{\mathbf{p}}}
   
   \def\bx{{\mathbf{x}}}
\def\by{{\mathbf{y}}}

\def\bI{{\mathbf{I}}}   
   
 \def\bR{{\mathbf{R}}}  
\def\bU{{\mathbf{U}}}

\begin{document}
\title{{Noncoherent Trellis Coded Quantization: A Practical Limited Feedback Technique for Massive MIMO Systems}}
\author{Junil Choi, Zachary Chance, David J. Love, and Upamanyu Madhow\\
\thanks{Junil Choi and David Love are with the School of Electrical and Computer Engineering, Purdue University, West Lafayette, IN (e-mail: choi215@purdue.edu, djlove@purdue.edu).}
\thanks{Zachary Chance is with MIT Lincoln Laboratory, Lexington, MA (e-mail: zac.chance@gmail.com).}
\thanks{Upamanyu Madhow is with the Department of Electrical and Computer Engineering, University of California, Santa Barbara, CA (e-mail: madhow@ece.ucsb.edu).}
\thanks{This paper was presented in part at the Information Theory and Application workshop, UCSD, 2013 \cite{ita_ntcq} and the IEEE Conference on Information Sciences and Systems, Johns Hopkins University, 2013 \cite{ciss_ntcq}.}
}\maketitle

\begin{abstract}
Accurate channel state information (CSI) is essential for attaining beamforming gains in single-user (SU) multiple-input multiple-output (MIMO) and multiplexing gains in multi-user (MU) MIMO wireless communication systems.  State-of-the-art limited feedback schemes, which rely on pre-defined codebooks for channel quantization, are only appropriate for a small number of transmit antennas and low feedback overhead.  In order to scale informed transmitter schemes to emerging massive MIMO systems with a large number of transmit antennas at the base station, one common approach is to employ time division duplexing (TDD) and to exploit the implicit feedback obtained from channel reciprocity. However, most existing cellular deployments are based on frequency division duplexing (FDD), hence it is of great interest to explore backwards compatible massive MIMO upgrades of such systems.  For a fixed feedback rate per antenna, the number of codewords for quantizing the channel grows exponentially with the number of antennas, hence generating feedback based on look-up from a standard vector quantized codebook does not scale. In this paper, we propose noncoherent trellis-coded quantization (NTCQ), whose encoding complexity scales linearly with the number of antennas. The approach exploits the duality between source encoding in a Grassmannian manifold (for finding a vector in the codebook which maximizes beamforming gain) and noncoherent sequence detection (for maximum likelihood
decoding subject to uncertainty in the channel gain). Furthermore, since noncoherent detection can be realized near-optimally using a bank of coherent detectors, we obtain a low-complexity implementation of NTCQ encoding using an off-the-shelf Viterbi algorithm applied to standard trellis coded quantization. We also develop advanced NTCQ schemes which utilize various channel properties such as temporal/spatial correlations. Monte Carlo simulation results show the proposed NTCQ and its extensions can achieve near-optimal performance with moderate complexity and feedback overhead.
\end{abstract}


\section{Introduction}\label{sec1}
\PARstart{T}{he} concept of wireless systems employing a large number of transmit antennas, often dubbed massive multiple-input multiple-output (MIMO) systems, has been evolving over the past few years.  It was found in \cite{massive_mimo3} that adding more antennas at the base station is always beneficial even with very noisy channel estimation because the base station can recover information even with a low signal-to-noise-ratio (SNR) once it has sufficiently many antennas.  This motivates the concept of using a very large number of transmit antennas, where the number of antenna elements can be at least an order of magnitude more than the current cellular systems (10s-100s) \cite{massive_mimo1}.  Massive MIMO systems have the potential to revolutionize cellular deployments by accommodating a large number of users in the same time-frequency slot to boost the network capacity \cite{massive_mimo2} and by increasing the range of transmission with improved power efficiency \cite{massive_mimo4}. Recently, fundamental limits, optimal transmit precoding and receive strategies, and real channel measurement issues for massive MIMO systems were studied and summarized in \cite{massive_mimo5} (see also the references therein).

When the transmitter has multiple antennas, channel state information (CSI) can provide significant performance gains, including beamforming gains in single-user (SU) multiple-input multiple-output (MIMO) systems and multiplexing gains in multi-user (MU) MIMO systems.  Unlike conventional MU-MIMO systems with a small number of transmit antennas, massive MU-MIMO can be implemented with simple per-user beamforming such as matched beamforming due to the large number of degrees-of-freedom available in the user channels \cite{massive_mimo1}.  However, without accurate CSI, massive MU-MIMO systems would also experience a sum-rate saturation, which is known as a \textit{ceiling effect}, even if the base station transmit power is unconstrained \cite{mumimo1,mumimo2}.

The challenge, therefore, is to scale channel estimation and feedback strategies to effectively provide CSI.  Most of the literature on massive MIMO sidesteps this challenge by focusing on  time division duplexing (TDD), for which CSI can be extracted {\it implicitly} using reciprocity.  However, since most cellular systems today employ frequency division duplexing (FDD), it is of great interest to explore effective approaches for obtaining CSI for massive MIMO upgrades of such systems. This motivates the work in this paper, which explores efficient approaches for quantizing high-dimensional channel vectors to generate CSI feedback.

There is a large body of literature devoted to accurate CSI quantization for closed-loop MIMO FDD systems with a relatively small number of antennas \cite{jsac_limited_feedback}.  Most approaches employ a common vector quantized (VQ) codebook at the transmitter and the receiver, and the explicit feedback sequence is simply the binary index of the codeword chosen in the codebook.  Thus, the main focus has been on codebook design.  For i.i.d. Rayleigh fading channel models, deterministic codebook techniques using Grassmannian line packing (GLP) were developed in \cite{grass1,grass2,grass3}, and the performance of random vector quantization (RVQ) codebooks was analyzed in \cite{rvq2,rvq3}.  Limited feedback codebooks that adapt to spatially correlated channels were studied in \cite{sp_correlated0,sp_correlated1,sp_correlated2}, and temporal correlated channels were developed in \cite{tm_correlated1,tm_correlated2,tm_correlated7,tm_correlated8,tm_correlated3,tm_correlated4,tm_correlated5,tm_correlated6}.

It has been shown in \cite{rvq2} that an RVQ codebook is asymptotically optimal for i.i.d. Rayleigh fading channels when the number of transmit antennas gets large, assuming a fixed number of feedback bits per antenna.
However, existing codebook-based techniques do not scale to approach the RVQ benchmark. In order to maintain the same level of channel quantization error, the feedback overhead must increase proportional to the number of transmit antennas \cite{rvq1,rvq3}.  While the linear increase in feedback overhead with the number of antennas may be acceptable as we scale to massive MIMO, the corresponding exponential increase in codebook size makes a direct look-up approach for feedback generation infeasible.

In order to address this gap in source coding techniques, it is natural to turn to the duality between source and channel coding.  Just as RVQ provides a benchmark for source coding, random coding produces information-theoretic benchmarks for channel coding.  However, there are thousands of papers dedicated to practical channel code designs that aim to approach these benchmarks, with codes such as convolutional codes, Reed-Solomon codes, turbo codes, and LDPC codes implemented in practice \cite{lin}.  While these ideas can and have been leveraged for source coding, the measures of distortion used have been the Hamming or Euclidean distortion.  Our contribution in this paper is to establish and exploit the connection between source coding on the Grassmannian manifold (which is what is needed for the limited feedback application of interest to us) and channel coding for {\it noncoherent} communication.  We coin the term \textit{noncoherent trellis-coded quantization (NTCQ)} for the class of schemes that we propose and investigate.
Our approach avoids the computational bottleneck of look-up based codebooks, with encoding complexity scaling linearly with the number of antennas, and its performance is near-optimal, approaching that of RVQ.

{\bf Approach:}
Our NTCQ approach relies on two key observations:\\
(a) Quantization for beamforming requires finding a quantized vector, from among the available choices, that is best aligned with the true channel vector,
in terms of maximizing the magnitude of their normalized inner product. This corresponds to a search on the Grassmann manifold rather than in Euclidean space. We point out, as have others before us, that this source coding problem maps to a
channel coding problem of {\it noncoherent} sequence detection, where we try to find the most likely transmitted codeword subject to an unknown multiplicative
complex-valued channel gain. \\
(b) We know from prior work on noncoherent communication that a noncoherent block demodulator can be implemented near-optimally using a bank of
coherent demodulators, each with a different hypothesis on the unknown channel gain.  Furthermore, signal designs and codes for coherent
communication are optimal for noncoherent communication, as long as we adjust our encoding and decoding slightly to account for the ambiguity caused by the unknown channel gain.

The relationship between quantization based on a mean squared error cost function and channel coding for {\it coherent} communication over the AWGN channel has been exploited successfully in the design of trellis coded quantization (TCQ)\cite{tcq1}, in which the code symbols take values from a standard finite
constellation used for communication, such as phase shift keying (PSK) or quadrature amplitude modulation (QAM).  The quantized code vector can then
be found by using a Viterbi algorithm for trellis decoding.  Our observation (b) allows us to immediately extend this strategy to the noncoherent setting.  The code vectors for NTCQ can be exactly the same as in standard TCQ, but the encoder now consists of several Viterbi algorithms (in practice, a very small number) running in parallel, with a rule for choosing the best output.  Thus,
while approximating a beamforming vector on the Grassmann manifold as in (a) appears to be difficult, it can be easily solved by using several parallel searches in Euclidean space.  Furthermore, just as noncoherent channel codes inherit the good performance of the coherent codes they were constructed from, NTCQ inherits the good quantization performance of TCQ.

{\bf Contributions:} Our contributions are summarized as follows:\\
$\bullet$ We show that channel codes, and by analogy, source codes developed in a coherent setting can be effectively leveraged in the noncoherent
setting of interest in CSI generation for beamforming. As shown through both analysis and simulations, the resulting NTCQ strategy provides near-optimal beamforming gain, and has encoding complexity which is linear in the channel dimension. \\
$\bullet$ We also develop adaptive NTCQ techniques that are optimized for spatial and temporal correlations.  A differential version of NTCQ utilizes the temporal correlation of the channel to successively refine the quantized channel to decrease the quantization error.  A spatially adaptive version of NTCQ exploits the spatial correlation of the channel so that it only quantizes the local area of the dominant direction of the spatial correlation matrix.
Utilization of channel statistics using such advanced schemes can significantly improve the performance or decrease the feedback overhead by utilizing channel statistics.

An important feature of NTCQ is its flexibility, which makes it an attractive candidate for potentially providing a common
channel quantization approach for heterogeneous fifth generation (5G) wireless communication systems, which could
involve a mix of advanced network entities such as massive MIMO, coordinated multipoint (CoMP) transmission, relay, distributed antenna systems (DAS), and femto/pico cells.
For example, massive MIMO systems could be implemented using a two-dimensional (2D) planar antenna array at the base station to reduce the size of antenna array \cite{fdmimo}.  Depending on the channel quality, the base station could turn on and off the rows/columns of this 2D array to achieve better performance.  The same situation could be encountered in CoMP and DAS because the number of coordinating transmit stations may vary over time.  NTCQ can easily adjust to such scenarios, since it can adapt to different numbers
of transmit antennas (or more generally, space-time channel dimension) by changing the number of code symbols, and can adapt CSI accuracy and feedback overhead by changing the constellation size and the coded modulation scheme.

{\bf Related work:} We have already mentioned conventional look-up based quantization approaches and discussed why they do not scale.  Trellis-based quantizers for CSI generation have been proposed previously in \cite{CK,ntcq1,ntcq2,Mingguang}, but the path metrics used for the trellis search are {\it ad hoc.} On the other hand, the mapping to noncoherent sequence detection, similar to NTCQ, has been pointed out in \cite{QAM}.  Depending on the number of constellation points used for the candidate codewords, the proposed algorithms in \cite{QAM} are dubbed as PSK \& QAM singular vector quantization (SVQ).  Although PSK/QAM-SVQ adopt similar codeword search methods as NTCQ, they do not consider coding.  The use of nontrivial trellis codes as proposed here significantly enhances performance compared to PSK/QAM-SVQ with the same amount of feedback overhead.  Furthermore, \cite{QAM} employs optimal noncoherent block demodulation, derived in \cite{qam_complexity,Sweldens}, for quantization,  incurring complexity $O(M_t^3)$ for QAM-SVQ and $O(M_t\log M_t)$ for PSK-SVQ, where $M_t$ denotes the number of antennas. Our NTCQ scheme exhibits better complexity scaling: near-optimal demodulation in $O(M_t)$ complexity by running a small number of coherent decoders in parallel, as proposed in \cite{Madhow}, suffices for providing near-optimal quantization performance.

The remainder of this paper is organized as follows.  In Section \ref{sec2}, we describe the system model and fundamentals underlying NTCQ.  A detailed description of the NTCQ algorithm and its variation is provided in Section \ref{sec3}.  Advanced NTCQ schemes that exploit temporal and spatial correlation of channels are explained in Section \ref{advanced_NTCQ}.  In Section \ref{simul}, simulation results are presented, and conclusions follow in Section \ref{conclusion}.

\section{System Model and Theory}\label{sec2}
\subsection{System Setup}
\begin{figure*}[t]
  \centering
  \includegraphics[scale = 0.7]{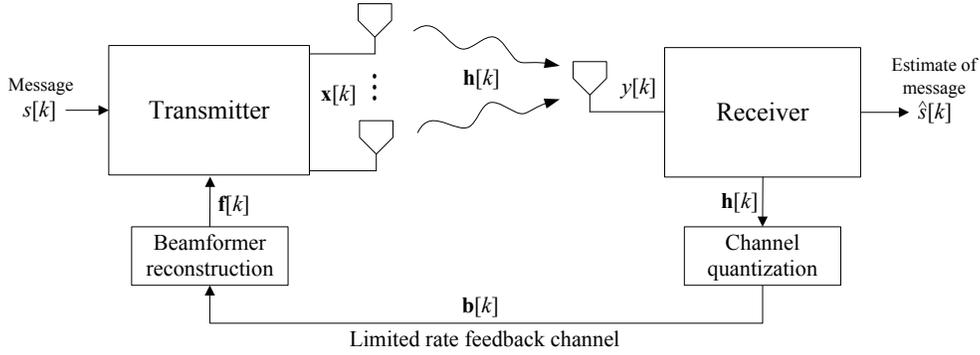}\\
  \caption{Multiple-input, single-output communications system with feedback.}\label{MISO}
\end{figure*}

We consider a block fading multiple-input single-output (MISO) communications system with $M_t$ transmit antennas at the transmitter as in Fig. \ref{MISO}.  The received signal, $y_\ell[k] \in \mathbb{C}$, for a channel use index $\ell$ in the $k$th fading block can be written as\footnote{Lower- and upper-case bold symbols denote vectors and matrices, respectively. The two-norm of a
vector ${\bx}$ is denoted as $\|\bx\|_2$. The transpose and Hermitian transpose of a vector $\bx$ are denoted by
$\bx^T$, $\bx^H$ respectively.  The expectation operator is denoted by $E\left[ \cdot \right]$, and $X \sim {\cal CN}(m, \sigma^2)$ indicates that $X$ is a complex Gaussian random variable with mean $m$ and variance $\sigma^2$.}
\begin{equation*}
y_\ell[k] = \bh^H[k]\bff[k]s_\ell[k] + z_\ell[k],
\end{equation*}
where $\bh[k] \in \mathbb{C}^{M_t}$ is the MISO channel vector, $\bff[k] \in \mathbb{C}^{M_t}$ is the beamforming vector with $\|\bff[k]\|^2_2=1$, $s_\ell[k] \in \mathbb{C}$ is the message signal with $E\left[s_\ell\left[k\right]\right] = 0$ and $E\left[|s_\ell[k]|^2\right] = \rho$, and $z_\ell[k] \in \mathbb{C}$ is additive complex Gaussian noise such that $z_\ell[k] \sim \cC\cN(0,\sigma^2)$. A number of different models for
$\bh[k]$ will be considered in the design and performance evaluation of quantization schemes, but for now, we allow it to be arbitrary.  The receiver quantizes its estimate of $\bh[k]$ into a $B_{\mathrm{tot}}$-dimensional binary vector $\bb [k]$, which is sent over a limited rate feedback channel.
The transmitter uses this feedback to construct a beamforming vector $\bff [k]$.  In order to focus attention on channel quantization, we do not model channel estimation errors at the receiver or errors over the feedback channel.

Since we do not consider temporal correlation in $\{ \bh [k] \}$ for quantizer design in this section, we drop the time index $k$ for the remainder of
this section. Assuming an average power constraint at the transmitter, we wish to choose $\bff$ so as to maximize the {\it normalized beamforming gain} that is defined as
\begin{equation} \label{beamforming_gain}
J ( \bff , \bh ) = \frac{|\bh^H\bff|^2}{{\|\bh\|_2^2}{\|\bff\|_2^2}}.
\end{equation}
Although $\|\bff\|_2=1$, we still normalize with $\|\bff\|_2$ in \eqref{beamforming_gain} to maintain notational generality.  
An equivalent approach is to minimize the {\it chordal distance} between $\bff$ and $\bh$, defined as
\begin{equation*} 
d_c^2 ( \bff , \bh ) = 1 - J ( \bff , \bh ) = 1 -  \frac{|\bh^H\bff|^2}{{\|\bh\|_2^2}{\|\bff\|_2^2}}.
\end{equation*}
These performance measures require searching for codewords on the Grassmann manifold, a projective space in which vectors are mapped to one-dimensional complex subspaces.

Conventional VQ codebook-based channel quantization typically employs
exhaustive search to select a codeword from  an unstructured and fixed $B_{\mathrm{tot}}$-bit codebook $\cC = \{\bc_{1},\bc_{2},\ldots,\bc_{2^{B_{\mathrm{tot}}}}\}$ according to
\begin{equation} \label{vec1}
\bc_{\mathrm{opt}} = \argmax_{\bc \in \cC} J ( \bc , \bh ) = \argmin_{\bc \in \cC} d_c^2 ( \bc , \bh ),
\end{equation}
and the binary sequence $\bb=\mathrm{bin}(\mathrm{opt})$ is fed back to the transmitter where $\mathrm{bin}(\cdot)$ converts an integer to its binary representation.  Then the beamforming vector is reconstructed at the transmitter as
\begin{equation*}
\bff=\frac{\bc_{\mathrm{int}(\bb)}}{\|\bc_{\mathrm{int}(\bb)}\|_2}
\end{equation*}
where $\mathrm{int}(\cdot)$ converts a binary string into an integer.  Exhaustive search, which does not require geometric interpretation of the performance metric,
incurs computational complexity $O(M_t2^{B_{\mathrm{tot}}})$, which is exponential in the number of bits.
We shall see that utilizing the geometry of the Grassmann manifold, and in particular, relating it to Euclidean geometry, is key to more efficient quantization procedures.

Since our performance criterion is independent of the codeword norm, one could, without loss of generality, normalize the codewords to unit norm up front (i.e., set $\|\bc\|_2 \equiv 1$).  However, for the code constructions and quantizer designs of interest to us, it is useful to allow codewords to have different norms (the performance criterion, of course, remains independent of codeword scaling).

\subsection{Feedback Overhead}
The relation between the feedback overhead $B_{\mathrm{tot}}$ (or codebook size $2^{B_{\mathrm{tot}}}$) and the performance of MIMO systems has been thoroughly investigated for i.i.d. Rayleigh fading channels.  In single user (SU) MISO channels with the $B_{\mathrm{tot}}$ bits RVQ codebook, the loss in normalized beamforming gain is given as \cite{rvq3}
\begin{align}\label{rvq}
\nonumber E\left[1-\max_{\bff\in \cF_{\mathrm{RVQ}}} J ( \bff , \bh )\right]&=2^{B_{\mathrm{tot}}} \beta \left(2^{B_{\mathrm{tot}}},\frac{M_t}{M_t-1}\right)\\
&\approx2^{-\frac{B_{\mathrm{tot}}}{M_t-1}}
\end{align}
where $\cF_{\mathrm{RVQ}}$ is an RVQ codebook, $\beta (x,y)=\frac{\Gamma(x)\Gamma(y)}{\Gamma(x+y)}$ is the Beta function, $\Gamma(x)=\int_0^{\infty} t^{x-1}e^{-t}dt$ is the Gamma function, and expectation is taken over $\bh$ and $\cF_{\mathrm{RVQ}}$.  The expression in \eqref{rvq} indicates that the feedback overhead needs to be increased proportional to $M_t$ to maintain the loss in normalized beamforming gain at a certain level.

For MU-MIMO zero-forcing beamforming (ZFBF), a similar conclusion is drawn in \cite{mumimo1,mumimo2}:
in order to achieve the full multiplexing gain of $M_t$,
the number of feedback bits per user, $B_{\mathrm{user}}$, must scale linearly with SNR (in dB) and $M_t$ as
\begin{equation*}
    B_{\mathrm{user}}=(M_t-1)\log_2\rho\approx \frac{M_t-1}{3}\rho_{dB}.
\end{equation*}

We therefore assume that at each channel use, the receiver sends back a binary feedback sequence of length
\begin{align*}
B_{\mathrm{tot}} &\triangleq B M_t+q
\end{align*}
where $B$ is the number of quantization bits used per transmit antenna and $q$ is a small, fixed number of auxiliary feedback bits, which does not scale with $M_t$.

While linear scaling of feedback bits with the number of transmit elements is typically acceptable in terms of overhead,
a VQ codebook-based limited feedback is computationally infeasible for massive MIMO systems with large $M_t$ because of the exponential growth of codeword search complexity with $M_t$ as $O(M_t 2^{BM_t})$.  Thus, we need to develop new techniques to quantize CSI for large $M_t$.

In order to develop an efficient CSI quantization method for massive MIMO systems,
we draw an analogy between searching for a candidate beamforming vector to maximize beamforming gain as in \eqref{vec1} and noncoherent sequence detection (e.g., \cite{QAM,CK}).  We then employ prior work relating noncoherent and coherent detection to map quantization
on the Grassmann manifold to quantization in Euclidean space, which can be accomplished far more efficiently.
This line of reasoning, which corresponds to the {\it process} of quantization, has been previously established in \cite{QAM}, but we provide
a self-contained derivation in Section \ref{sec:encoding} pointing to a low-complexity, near-optimal source encoding strategy.
We then show, in Section \ref{sec:euclidean_codebooks} that structured quantization codebooks for Euclidean metrics are effective for quantization
on the Grassmann manifold.  This leads to a CSI quantization framework which is efficient in terms of both overhead and computation.

\subsection{Efficient Grassmannian Encoding using Euclidean Metrics} \label{sec:encoding}

Consider a single antenna noncoherent, block fading, additive white Gaussian noise (AWGN) channel with received vector
\begin{equation*}
\by = \beta \bx + \bn,
\end{equation*}
\noindent where $\beta \in \mathbb{C}$ is an unknown complex channel gain, $\bx \in \mathbb{C}^{N}$ is a vector of $N$ transmitted symbols, $\bn \in \mathbb{C}^{N}$ is complex Gaussian noise, and $\by \in \mathbb{C}^{N}$ is the received signal.  Using the generalized likelihood ratio test (GLRT) as in \cite{QAM,Madhow}, the estimate of the transmitted vector, $\hat{\bx}$, is given by
\begin{align}
\hat{\bx} &= \argmin_{\bx\in \mathbb{C}^{N}}\min_{\beta\in \mathbb{C}}\|\by - \beta\bx\|_{2}^2\label{nonco}\\
&=\argmin_{\bx\in \mathbb{C}^{N}}\min_{\alpha\in \mathbb{R}^+}\min_{\theta\in [0,2\pi)}\|\by\|_{2}^2 + \alpha^2\|e^{j\theta}\bx\|_{2}^2 - 2\alpha\Re(e^{j\theta}\by^H\bx)\label{decompose}\\
&=\argmin_{\bx\in \mathbb{C}^{N}}\min_{\alpha\in \mathbb{R}^+}\|\by\|_{2}^2 + \alpha^2\|\bx\|_{2}^2 - 2\alpha|\by^H\bx|\label{differentiate}\\
&=\argmax_{\bx\in \mathbb{C}^{N}}\frac{|\by^H\bx|^2}{\|\bx\|_2^2},\label{equiv1}
\end{align}
\noindent where we decomposed the entire complex plain $\beta=\alpha e^{j\theta}$ with $\alpha\in \mathbb{R}^+$ and $\theta \in [0,2\pi)$ in \eqref{decompose}, and \eqref{differentiate} comes from
\[
\min_{\theta\in [0,2\pi)}\left\{-\Re(e^{j\theta}\by^H\bx)\right\}=-|\by^H\bx|.
\]
To derive \eqref{equiv1}, we differentiate \eqref{differentiate} with respect to $\alpha$ and set to $0$ which gives $\alpha^\star=\frac{|\by^H\bx|}{\|\bx\|_2^2}$.  Note that $\alpha^\star$ is the global minimizer of \eqref{differentiate} because \eqref{differentiate} is a quadratic function of $\alpha$.  We can derive \eqref{equiv1} after plugging $\alpha^\star$ into \eqref{differentiate} and some basic algebra.

We can easily check from \eqref{vec1} and \eqref{equiv1} that finding the optimal codeword for a MISO beamforming system and the noncoherent sequence detection  problems are equivalent (although this relation is already shown in \cite{QAM}, we proved the duality of \eqref{nonco} and \eqref{vec1} more explicitly than \cite{QAM}).  Therefore, we can find $\bc_{\mathrm{opt}}$ for a MISO beamforming system with a Euclidean distance quantizer (or noncoherent block demodulator)
\begin{align}\label{vec3}
\min_{\alpha \in \mathbb{R}^+}\min_{\theta \in [0,2\pi)}\min_{\bc_i\in \cC}\|\bar{\bh} - \alpha e^{j\theta}\bc_i\|_{2}^2.
\end{align}
where $\bar{\bh}=\frac{\bh}{\|\bh\|_2}$ is the normalized channel direction.

Moreover, instead of searching over the entire complex plane by having $\alpha \in \mathbb{R}^+$ and $\theta\in [0,2\pi)$, we know from prior work on noncoherent communication \cite{Madhow} that the noncoherent block demodulator in \eqref{vec3} can be implemented near-optimally using a bank of coherent demodulators over the optimized discrete sets of $\alpha \in \mathbb{A} = \{\alpha_{1},\alpha_{2},\ldots,\alpha_{K_{\alpha}}\}$ and $\theta \in \Theta = \{\theta_{1},\theta_{2},\ldots,\theta_{K_{\theta}}\}$.  While {\it optimal} noncoherent detection can be accomplished
with quadratic complexity in $M_t$ \cite{QAM}, as we show through our numerical results, a small number of parallel coherent demodulators
(which incurs complexity linear in $M_t$) is all that is required for excellent quantization performance.

The preceding development tells us that we can apply coherent demodulation, which maps to quantization using Euclidean metrics, to noncoherent demodulation, which maps to quantization on the Grassmann manifold.  However, we must still determine how to choose the quantization codebook.  Next, we present results indicating that we can simply use codes optimized for Euclidean metrics for this purpose.

\subsection{Efficient Grassmannian Codebooks based on Euclidean Metrics} \label{sec:euclidean_codebooks}

We begin with an asymptotic result for i.i.d. Rayleigh fading coefficients, which relies on the well-known rate-distortion theory for i.i.d. Gaussian sources.
\begin{theorem}\label{l_theorem}
If we quantize an $M_t \times 1$ i.i.d. Rayleigh fading MISO channel $\bh \sim\mathcal{CN}(\mathbf{0},\sigma_h^2\bI)$ with a Euclidean distance quantizer using $B$ bits per entry (which corresponds to $\frac{B}{2}$ bits per each of real and imaginary dimension) as
\begin{equation}\label{ED_quantizer}
\bg_{\mathrm{ED}}=\min_{\bg_i \in \mathcal{G}}\|\bh-\bg_i\|_2^2
\end{equation}
where $\mathcal{G}=\left\{\bg_1,\ldots,\bg_{2^{B_{\mathrm{tot}}}}\right\}$, $B_{\mathrm{tot}}=BM_t$, $\bg_i\sim \mathcal{CN}(\mathbf{0},(\sigma_h^2-2D)\bI)$ for all $i$, and $D=\frac{1}{2}\sigma_h^22^{-B}$, then the asymptotic loss in normalized beamforming gain, or chordal distance, is given by
\begin{equation}\label{g_loss}
d_c^2 ( \bh, \bg_{\mathrm{ED}} ) \stackrel{M_t \rightarrow \infty}{\longrightarrow}2^{-B}.
\end{equation}
\end{theorem}
\begin{IEEEproof}
By expanding $\|\bh-\bg_{\mathrm{ED}}\|_2^2$, we have

\begin{align*}
\|\bh-\bg_{\mathrm{ED}}\|_2^2=\sum_{t=1}^{M_t}&\left[\left\{\mathrm{Re}(h_t)-\mathrm{Re}(g_{\mathrm{ED},t})\right\}^2\right.\\
&\qquad\left.+\left\{\mathrm{Im}(h_t)-\mathrm{Im}(g_{\mathrm{ED},t})\right\}^2\right]
\end{align*}
where $h_t$ and $g_{\mathrm{ED},t}$ are the $t^{th}$ entry of $\bh$ and $\bg_{\mathrm{ED}}$, respectively.  Note that $\mathrm{Re}(h_t)$ and $\mathrm{Im}(h_t)$ are from the same distribution $\mathcal{N}(0,\frac{1}{2}\sigma_h^2)$, and $\mathrm{Re}(g_{\mathrm{ED},t})$ and $\mathrm{Im}(g_{\mathrm{ED},t})$ are from the distribution $\mathcal{N}(0,\frac{1}{2}\sigma_h^2-D)$.  Assuming $\frac{B}{2}$ bits are used to quantize each of $\mathrm{Re}(h_t)$ and $\mathrm{Im}(h_t)$ for all $t$, by rate-distortion theory for i.i.d. Gaussian sources \cite{Gallager}, we can achieve the rate-distortion bound
\begin{align*}
E\left[\left\{\mathrm{Re}(h_t)-\mathrm{Re}(g_{\mathrm{ED},t})\right\}^2\right]&=E\left[\left\{\mathrm{Im}(h_t)-\mathrm{Im}(g_{\mathrm{ED},t})\right\}^2\right]\\
&=D
\end{align*}
as $M_t \rightarrow \infty$.  Thus, by the weak law of large numbers, the following convergences hold\footnote{Let $\bar{X}_n=\frac{1}{n}(X_1+\cdots+X_n)$ and $\mu=E[X_i]$ for all $i$.  We say $\bar{X}_n$ converges to $\mu$ in probability as $\bar{X}_n\stackrel{P}{\rightarrow}\mu$ for $n\rightarrow \infty$ when $\lim\limits_{n\rightarrow \infty}Pr\left(|\bar{X}_n-\mu|>\epsilon\right)=0$ for any $\epsilon>0$.}
\begin{align*}
\frac{1}{M_t}  \|\bh-\bg_{\mathrm{ED}}\|_2^2&\stackrel{P}{\rightarrow} 2E\left[\left\{\mathrm{Re}(h_t)-\mathrm{Re}(g_{\mathrm{ED},t})\right\}^2\right]=2D,\\
\frac{1}{M_t}  \|\bh\|_2^2&\stackrel{P}{\rightarrow} 2E[\left\{\mathrm{Re}(h_t)\right\}^2]=\sigma_h^2,\\
\frac{1}{M_t}  \|\bg_{\mathrm{ED}}\|_2^2&\stackrel{P}{\rightarrow} 2E[\left\{\mathrm{Re}(g_{\mathrm{ED},t})\right\}^2]=\sigma_h^2-2D
\end{align*}
as $M_t\rightarrow \infty$.  Moreover, $\left|\frac{\bh^H\bg_{\mathrm{ED}}}{M_t}\right|^2$ can be lower bounded as
\begin{align*}
  \left|\frac{\bh^H\bg_{\mathrm{ED}}}{M_t}\right|^2 &\geq \left(\frac{\mathrm{Re}(\bh^H\bg_{\mathrm{ED}})}{M_t}\right)^2\\
  &=\left(\frac{\|\bh\|_2^2+\|\bg_{\mathrm{ED}}\|_2^2-\|\bh-\bg_{\mathrm{ED}}\|_2^2}{2M_t}\right)^2\\
  &\stackrel{P}{\rightarrow}\left(\sigma_h^2-2D\right)^2.
\end{align*}
Then, the normalized beamforming gain loss relative to the unquantized beamforming case is bounded as
\begin{align*}
  d_c^2 ( \bh, \bg_{\mathrm{ED}} )&=1-\frac{|\bh^H\bg_{\mathrm{ED}}|^2}{\|\bh\|_2^2\|\bg_{\mathrm{ED}}\|_2^2}
  \leq \frac{2D}{\sigma_h^2}=2^{-B},\\
  d_c^2 ( \bh, \bg_{\mathrm{ED}} )&\stackrel{(a)}{\geq} 2^{-\frac{BM_t}{M_t-1}}
\end{align*}
where $(a)$ follows from the optimality of the RVQ codebook in large asymptotic regime \cite{rvq2}.  As $M_t\rightarrow \infty$, the lower bound of $d_c^2 ( \bh, \bg_{\mathrm{ED}} )$ converges to the upper bound $2^{-B}$, which finishes the proof.
\end{IEEEproof}

Note that the loss in \eqref{g_loss} is asymptotically the same as that of the RVQ codebook in \eqref{rvq}.  Since the RVQ codebook is
known to be asymptotically optimal as $M_t \rightarrow \infty$ (fixing the number of bits per antenna) \cite{rvq2},
we conclude that coherent Euclidean distance quantization as in \eqref{ED_quantizer} with a rich, rotationally invariant constellation such as a Gaussian codebook $\mathcal{G}$, is also an asymptotically optimal way to quantize the channel vector $\bh$.
Of course, in practice, for finite constellations and number of antennas, we must ``align'' the codewords $\bg_i$ with the channel $\bh$,
using parallel branches with different amplitude scaling $\alpha$ and phase rotations $\theta$ as in \eqref{vec3}, prior to computing the Euclidean metric, in order to maximize the beamforming gain.

We also note that the use of nontrivial codes is implicit in Theorem \ref{l_theorem}, hence the uncoded constellations employed
in \cite{QAM} do not achieve optimal quantization performance.  The constellation expansion employed
in the NTCQ schemes considered here is required to approach optimal performance.

We now provide a {\it non-asymptotic} result regarding the
chordal distances associated with Grassmannian line packing (GLP) attained by
codebooks optimized using Euclidean metrics.
Let $N=2^{B_{\mathrm{tot}}}$ and $\mathcal{U}_{M_t}^{N}\in \mathbb{C}^{M_t \times N}$ denote the set of $M_t \times N$ complex matrices with unit vector columns.  To minimize the average quantization error of \eqref{vec3} or \eqref{ED_quantizer} in Euclidean space with a fixed codebook $\mathcal{C}$, we have to maximize the minimum Euclidean distance between all possible codeword pairs
\begin{equation*}
    d_{E,
    \min}^2(\mathcal{C})\triangleq \min_{1\leq k<l\leq N}d_E^2(\bc_k,\bc_l)
\end{equation*}
where $d_E(\bx,\by)\triangleq \|\bx-\by\|_2$, and $\{\bc_i\}_{i=1}^{N}$ are column vectors of $\mathcal{C}$.
Let $\mathcal{C}_{\mathrm{ED}}$ denote an optimized Euclidean distance (ED) codebook that maximizes the minimum Euclidean distance as
\begin{equation*}
  \mathcal{C}_{\mathrm{ED}}=\argmax_{\mathcal{C}\in\mathcal{U}_{M_t}^{N}}d^2_{E,\min}(\mathcal{C}).
\end{equation*}
On the other hand, beamforming codebooks are ideally designed for i.i.d. Rayleigh fading channels to maximize the minimum chordal distance between codewords as
\begin{equation*}
    d^2_{c,\min}(\mathcal{C})\triangleq \min_{1\leq k<l\leq N}d^2_c(\bc_k,\bc_l),
\end{equation*}
and a GLP codebook is given as \cite{grass1,grass3}
\begin{equation*}
    \mathcal{C}_{\mathrm{GLP}}=\argmax_{\mathcal{C}\in\mathcal{U}_{M_t}^{N}} d^2_{c,\mathrm{min}}(\mathcal{C}).
\end{equation*}
Note that the optimization metrics of $\mathcal{C}_{\mathrm{GLP}}$ and $\mathcal{C}_{\mathrm{ED}}$ are different, the former is the chordal  distance and the latter is the Euclidean distance.  The following lemma shows the relation of the two metrics.
\begin{lemma}\label{lemma1}
    For any two unit vectors $\bx$ and $\by$, the squared chordal distance between $\bx$ and $\by$ is upper bounded by a function of their Euclidean distance as
    \begin{align*}
      d^2_c(\bx,\by)&\leq1-\left(1-\frac{1}{2}d_E^2(\bx,\by)\right)^2\\
      &=d_E^2(\bx,\by)-\frac{1}{4}d_E^4(\bx,\by).
    \end{align*}
\end{lemma}
\begin{IEEEproof}
Let us define $d_{\theta}^2(\bx,\by)$ as
\begin{align*}
      d_{\theta}^2(\bx,\by)&\triangleq \min_{\theta\in[0,2\pi)}d_E^2(\bx,e^{j\theta}\by)\\
      &=\|\bx\|_2^2+\|\by\|_2^2-2\max_{\theta\in[0,2\pi)}\mathrm{Re}\left\{e^{j\theta}\bx^H\by\right\}\\
      &=2-2|\bx^H\by|\leq d_E^2(\bx,\by).
\end{align*}
Then, the squared chordal distance of $\bx$ and $\by$ is upper bounded as
\begin{align*}
    d^2_c(\bx,\by)&=1-|\bx^H \by|^2\\
    &=1-\left(1-\frac{1}{2}d_{\theta}^2(\bx,\by)\right)^2\\
    &\leq 1-\left(1-\frac{1}{2}d_E^2(\bx,\by)\right)^2,
\end{align*}
which finishes the proof.
\end{IEEEproof}

Moreover, Lemma \ref{lemma1} can be directly extended to the following corollary.
\begin{corollary}\label{corollary1}
The minimum chordal distance of $\mathcal{C}_{\mathrm{ED}}$, $d^2_{c,\mathrm{min}}(\mathcal{C}_{\mathrm{ED}})$, is upper bounded by the minimum Euclidean distance of $\mathcal{C}_{\mathrm{ED}}$, $d^2_{E,\min}(\mathcal{C}_{\mathrm{ED}})$ as
\begin{equation*}
  d^2_{c,\mathrm{min}}(\mathcal{C}_{\mathrm{ED}})\leq d^2_{E,\min}(\mathcal{C}_{\mathrm{ED}}).
\end{equation*}
\end{corollary}
\begin{figure}[t]
  \centering\includegraphics[width=1\columnwidth]{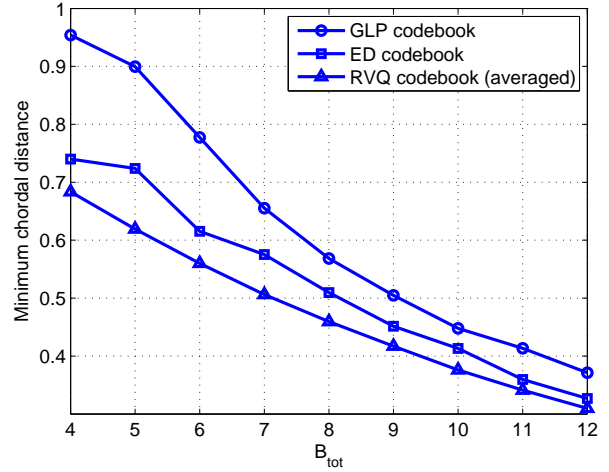}\\
  \caption{The minimum chordal distances of different codebooks with $M_t=8$.  GLP and Euclidean distance (ED) codebook are numerically optimized according to their metrics, while the minimum distance of RVQ codebook is averaged over 1000 different RVQ codebooks.}\label{dc_comparison}
\end{figure}
Although Corollary \ref{corollary1} does not say that $\mathcal{C}_{\mathrm{ED}}$ maximizes the minimum chordal distance between its codewords, $\mathcal{C}_{\mathrm{ED}}$ is expected to have a \textit{good} chordal distance property.  We verify this by simulation with numerically optimized $\mathcal{C}_{\mathrm{GLP}}$ and $\mathcal{C}_{\mathrm{ED}}$ in Fig. \ref{dc_comparison}.  It is shown that the minimum chordal distance of $\mathcal{C}_{\mathrm{ED}}$ is larger than the (averaged) minimum chordal distance of the RVQ codebook for all $B_{\mathrm{tot}}$ values.

\section{Noncoherent Trellis-Coded Quantization (NTCQ)}\label{sec3}
\subsection{Euclidean Distance Codebook Design}
The observations in the preceding section provide the following practical guidelines for quantization on the Grassmann manifold:
(a) find a good codebook in Euclidean space whose structure permits efficient encoding (or, equivalently, find a good, efficiently decodable channel code); (b) use parallel versions of the Euclidean encoder with different amplitude scalings and phase rotations, and choose the best output (or, equivalently, implement block noncoherent decoding efficiently with a number of parallel coherent decoders).  The proposed NTCQ emerges naturally from application of these guidelines.

NTCQ relies on trellis-coded quantization (TCQ) which was originally proposed in \cite{tcq1}, exploiting the functional duality between source coding and channel coding to leverage the well-known trellis-coded modulation (TCM) channel codes designed for coherent communication over AWGN channels \cite{Unger}.  TCM integrates the design of convolutional codes with modulation to maximize the minimum Euclidean distance between modulated codewords.  This is done by coding over partitions of the source constellation.  Let $\mathcal{C}_{\mathrm{TCM}}$ denote a fixed codebook with $N$ codewords generated by a TCM channel code.  Then $\mathcal{C}_{\mathrm{TCM}}$ can be mathematically expressed as
\begin{align*}
    \mathcal{C}_{\mathrm{TCM}}&=\argmax_{\mathcal{C}\in\mathcal{V}_{M_t}^{N}}d^2_{\mathrm{E,min}}(\mathcal{C})
\end{align*}
where $\mathcal{V}_{M_t}^{N}\subset \mathcal{U}_{M_t}^{N}$ is the set of $M_t \times N$ complex matrices generated by a given trellis structure with a finite number of constellation points of interest for entries of the matrix.  Note that $\mathcal{C}_{\mathrm{TCM}}$ is a Euclidean distance codebook within a given set $\mathcal{V}_{M_t}^{N}$.  Thus, $\mathcal{C}_{\mathrm{TCM}}$ is expected to have a \textit{good} chordal distance property as well.
\begin{figure}[t]
  \centering
  \includegraphics[scale = 0.5]{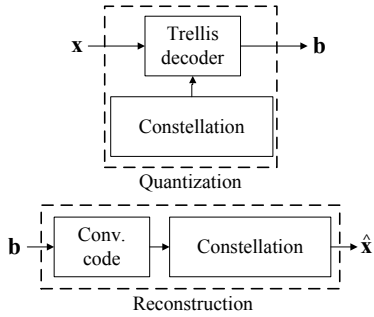}\\
  \caption{Quantization and reconstruction processes for a Euclidean distance quantizer using trellis-coded quantization (TCQ).}\label{CVQ}
\end{figure}

In TCQ, the decoder and encoder of TCM are used to quantize and reconstruct a given source, respectively.  From Fig. \ref{CVQ}, we see that the TCQ system consists of a source constellation, a trellis-based decoder (for source quantization), and a convolutional encoder (for source reconstruction).  Quantization is performed by passing a source vector $\bx\in \mathbb{C}^{N}$ through a trellis-based optimization whose goal is to minimize a mean square error distortion between the quantized output and the source message input.
The additive structure of the square of Euclidean distance implies that the Viterbi algorithm
can be employed to efficiently search for a codebook vector that minimizes the Euclidean distance from a given source vector as
\begin{equation}\label{tcq}
\bc_{\mathrm{opt}} = \argmin_{\bc_i\in\cC_{\mathrm{TCM}}}\|\bx - \bc_i\|^2_{2},
\end{equation}
which is then mapped to a binary sequence $\bb=\mathrm{bin}(\mathrm{opt})$.  The quantized source vector $\hat{\bx}$ is reconstructed by passing the binary sequence $\bb$ into the convolutional encoder and mapping the binary output of the convolutional encoder to points on the source constellation (as if modulating the signal).  Due to the linearity of the convolutional code, each unique binary sequence $\bb$ represents a unique quantized vector $\hat{\bx}$.

NTCQ adopts TCQ to quantize CSI.  Note that \eqref{tcq} is the same optimization problem as \eqref{vec3} with a given $\alpha \in \mathbb{A} = \{\alpha_{1},\alpha_{2},\ldots,\alpha_{K_{\alpha}}\}$ and $\theta \in \Theta = \{\theta_{1},\theta_{2},\ldots,\theta_{K_{\theta}}\}$.  Thus, the minimization \eqref{vec3} can be performed using $K_{\alpha}\cdot K_{\theta}$ parallel instances of the Viterbi algorithm.  This is the same paradigm proposed as in TCQ except for the search over $\alpha$ and $\theta$ parameters; due to the presence of these terms, the process is coined \textit{noncoherent trellis-coded quantization}.  Note that with PSK constellations, we can set $\alpha=1$ because all the candidate beamforming vectors $\bc_i$'s have the same norm.

We explain the implementation of NTCQ with 8PSK and 16QAM constellations next (we also report results for QPSK, but
do not describe the corresponding NTCQ procedure, since it is similar to that for 8PSK). Before explaining the actual implementation, it should be pointed out that, because of the inherited TCM structure, the number of constellation points is larger than $2^B$ in NTCQ where $B$ is the number of quantization bits per channel entry.  We explicitly list the relationship between $B$ and the constellations in Table \ref{bitperentry}.  This issue will become clear as we explain the 8PSK implementation.
\begin{table}
\caption{Mapping of quantizing bits/entry ($B$) and constellations.} \centering
\begin{tabular}{|c||c|c|c|}
  \hline
  $B$ & 1 bit/entry & 2 bits/entry & 3 bits/entry \\
  \hline
  Constellation & QPSK & 8PSK & 16QAM \\
  \hline
\end{tabular}
\label{bitperentry}
\end{table}

\subsection{NTCQ with 8PSK (2 bits/entry)}\label{8psk_explain}
\begin{figure}[t]
  \centering
  \includegraphics[scale = 0.55]{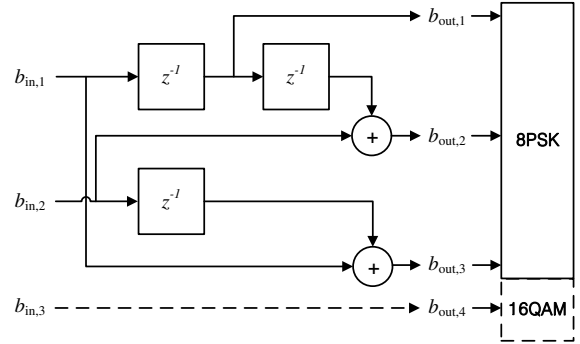}\\
  \caption{This rate 2/3 convolutional code corresponds to the trellis in Fig. \ref{trellis}.  In the figure, the smaller the index the less significant the bit, e.g., $b_{\mathrm{in},1}$ is the least significant input bit and $b_{\mathrm{in},3}$ is the most significant input bit.}\label{unger_8}
\end{figure}

\begin{figure}[t]
\centering
\includegraphics[scale = 1]{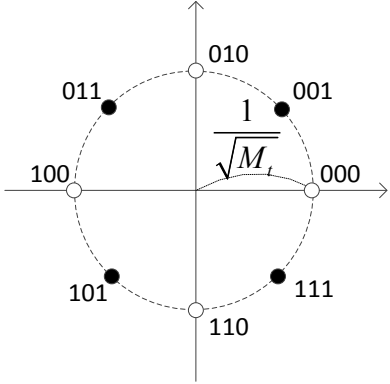}\\
\caption{8PSK constellation points used in NTCQ are labeled with binary sequences.}\label{psk}
\end{figure}

We adopt the rate 2/3 convolutional code in \cite{Unger}, as shown in Fig. \ref{unger_8}.  The source constellation is assumed to be 8PSK as in Fig. \ref{psk}.  Note that all constellation points are normalized with the number of transmit antennas $M_t$.

\begin{figure}[t]
  \centering
  \includegraphics[scale = 0.5]{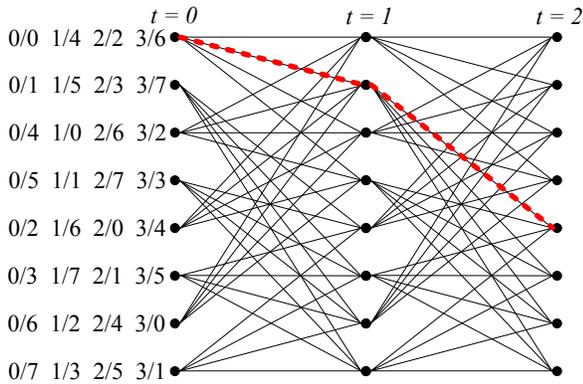}\\
  \caption{The Ungerboeck trellis with $S=8$ states corresponding to the convolutional encoder in Fig. \ref{unger_8}.  The input/output relations using decimal numbers correspond to state transitions from the top to bottom.  The example path $\bp_{2} = [1,2,5]$ that corresponds to binary input sequence $[01,00]^T$ (or decimal input $[1,0]^T$) and binary output sequence $[100,001]^T$ (or decimal output $[4,1]^T$) is highlighted.}\label{trellis}
\end{figure}

The construction of the feedback sequence is done using a trellis decoder.  As is done in traditional decoding of convolutional codes, the encoding process is represented using a trellis showing the relationship between states of the encoder along with input and output transitions.  The trellis with input/output state transitions corresponding to the convolutional code in Fig. \ref{unger_8} is shown in Fig. \ref{trellis}.

We select candidate beamforming vectors using an $M_t$-stage trellis where each stage selects an entry in each of the candidate vectors.  Thus, each path through the trellis corresponds to a unique candidate beamforming vector.  It is important to note that there are only four state-transitions from any of the eight states in Fig. \ref{trellis}.  Each transition is mapped to one point of the 8PSK constellation.  Therefore, even though the source constellation is 8PSK, each element of $\bar{\bh}$ is quantized with one of the QPSK subconstellations marked by black or white circles in Fig. \ref{psk}, which results in 2 bits quantization per entry as shown in Table \ref{bitperentry}.

The path choices are enumerated with binary labels, and each path also corresponds to a unique binary sequence.  The candidate vector or path that is chosen for output is the one that optimizes the given path metric.  The path metric is chosen to reflect the desired Euclidean distance minimization regarding codeword $\bc_i$ in (\ref{vec3}) for a given $\alpha$ and $\theta$.  The output of the quantization is the binary sequence corresponding to the best candidate path.

Each transition from each state at the $t^{th}$ stage, $s_{t} \in \{1, 2, \ldots, S\}$, in the trellis to a state at the $(t+1)^{th}$ stage, $s_{t+1}$, corresponds to a point in the source constellation.  For example, a transition from state 4 to state 8 corresponds to the binary output sequence 011 which corresponds to the constellation point $\frac{1}{\sqrt{2M_t}}~(-1 + j)$ in Fig. \ref{psk}.  Note that, in this setup, a single entry is chosen at each stage where it is possible to choose more; this is done by using intermediate codebooks for each stage of the trellis.  For more details on this method and the design of the codebooks, the reader is referred to \cite{CK}.

To optimize over the trellis, the first task is to define a path metric.  Let $\bp_{t}$ be a partial path, or a sequence of states, up to the stage $t$.  For example, the path $\bp_{2} = [1,2,5]$ using state indices is highlighted in Fig. \ref{trellis}.  Also, define the two functions $\inf(\cdot)$ and $\outf(\cdot)$ such that $\inf(\bp_t)$ outputs the binary input sequence corresponding to path $\bp_{t}$, and $\outf(\bp_{t})$ gives the sequence of output constellation points corresponding to the path $\bp_{t}$.  Again, using the sample path $\bp_{2}$ in Fig. \ref{trellis}, we can see that
\begin{equation*}
\inf(\bp_{2}) = [01,00]^T,~\outf(\bp_{2}) = \frac{1}{\sqrt{M_t}}\left[-1,\frac{1}{\sqrt{2}}(1+j)\right]^T.
\end{equation*}

With these definitions, we can define the path metric, $m(\cdot)$, as
\begin{equation*}
m(\bp_{t},\theta) = \|\bar{\bh}_{t} - e^{j\theta}\outf(\bp_{t})\|_2^2,
\end{equation*}
where $\theta\in [0,2\pi)$ and $\bar{\bh}_{t}$ is the vector created by truncating of normalized MISO channel vector $\bar{\bh}$ to the first $t$ entries.  Note that $\alpha=1$ because all constellation points have the same magnitude in the 8PSK case.  It is easy to check that minimizing over the path metric will minimize the Euclidean distance.  It is also important to notice that the path metric can be written recursively as
\begin{equation*}
m(\bp_{t},\theta) = m(\bp_{t-1},\theta) + \left|\bar{h}_{t} - e^{j\theta}\outf\left([p_{t-1}~p_{t}]^T\right)\right|^2,
\end{equation*}
where $\bar{h}_{t}$ and $p_t$ are the $t^{th}$ entry of $\bar{\bh}$ and $\bp_t$, respectively.  The above path metric can be efficiently computed via the Viterbi algorithm.  The path metric is computed in parallel for each quantized value of $\theta \in \Theta = \{\theta_{1},\theta_{2},\ldots,\theta_{K_{\theta}}\}$. Then the best path $\bp_{\mathrm{best}}$ and the phase $\theta_{\mathrm{best}}$ that minimize the path metric can be found as
\begin{equation*}
\min_{\theta\in \Theta}\min_{\bp_{M_t}\in\mathbb{P}_{M_t}}m(\bp_{M_t},\theta)\label{pbest}
\end{equation*}
where $\mathbb{P}_{M_t}$ denotes all possible paths up to stage $M_t$.  Finally, the beamforming vector $\bff$ is calculated as
\begin{equation*}
\bc_{\mathrm{opt}}=\outf(\bp_{\mathrm{best}}),~\bff=\frac{\bc_{\mathrm{opt}}}{\|\bc_{\mathrm{opt}}\|_2}.
\end{equation*}
Note that $\|\bc_{\mathrm{opt}}\|_2=1$ for 8PSK; therefore $\bff=\bc_{\mathrm{opt}}$.

It is important to point out that minimizing over $\theta$ only increases the complexity of quantization, not the feedback overhead because the transmitter does not have to know the value of $\theta_{\mathrm{best}}$ that minimizes the path metric during the beamforming vector reconstruction process.  However, there is additional feedback overhead with NTCQ.  Since we test all paths in the trellis, the transmitter has to know the starting state of $\bp_{\mathrm{best}}$, which causes additional $\log_2S$ bits of feedback overhead where $S$ is the number of states in the trellis.  Therefore, the total feedback overhead is
\begin{equation*}
  B_{\mathrm{tot}}=BM_t+\log_2S.
\end{equation*}
The additional feedback overhead $\log_2S$ bits can vary depending on the trellis used in NTCQ.

\subsection{NTCQ with 16QAM (3 bits/entry)}\label{16qam_explain}
\begin{figure}[t]
\centering
\includegraphics[scale = 1]{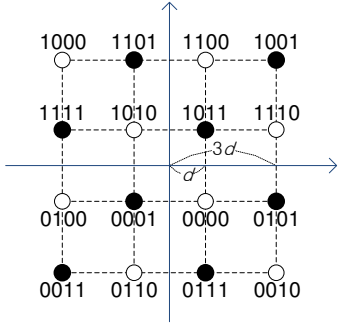}\\
\caption{16QAM constellation points used in NTCQ are labeled with binary sequences.}\label{16QAM}
\end{figure}
For the 16QAM constellation, the rate 3/4 convolution encoder is shown in Fig. \ref{unger_8}.  The source constellation is shown in Fig. \ref{16QAM} where $d=\frac{\triangle}{2\sqrt{M_t}}$ with $\triangle=\sqrt{\frac{6}{M-1}}$ with $M=16$ to have $E[\|\bc_i\|_2^2]=1$ where expectation is taken over $\bc_i$ assuming all constellation points are selected with equal probability.

The procedure of NTCQ using 16QAM is basically the same as the 8PSK case. The difference arising for 16QAM is that we have to take $\alpha$ into account during the path metric computation as
\begin{equation}\label{16qam_optimizing}
m(\bp_{t},\alpha,\theta) = \|\bar{\bh}_{t} - \alpha e^{j\theta}\outf(\bp_{t})\|_2^2
\end{equation}
where $\theta \in \Theta = \{\theta_{1},\theta_{2},\ldots,\theta_{K_{\theta}}\}$ and $\alpha \in \mathbb{A} = \{\alpha_{1},\alpha_{2},\ldots,\alpha_{K_{\alpha}}\}$.  Similar to the 8PSK case, additional $\log_2S$ feedback bits are needed to indicate the starting state of $\bp_{\mathrm{best}}$ to the transmitter in the 16QAM case.

\subsection{Complexity}\label{complexity_analysis}
NTCQ relies on a trellis search to quantize the beamforming vector, and the trellis search is performed by the Viterbi algorithm.  In each state transition of the trellis, one channel entry is quantized with one of $2^B$ constellation points.  This computation is performed for $S$ states in each state transition (stage) and there are $M_t$ state transitions in total.  Thus, the complexity of the Viterbi algorithm becomes $O(2^BSM_t)$.

The Viterbi algorithm has to be executed $K_{\theta}\cdot K_{\alpha}$ times in NCTQ, which gives the overall complexity of $O(K_{\theta} K_{\alpha}2^BSM_t)$.  In the limit of large $M_t$, Theorem \ref{l_theorem} tells us that we can get away with $K_{\theta}\rightarrow 1$ and $K_{\alpha}\rightarrow 1$
without performance loss.  However, even for moderate values of $M_t$, our results in Section \ref{iid}
show that small values of $K_{\theta}$ and $K_{\alpha}$ can be employed with minimal performance degradation.
The key aspect to note is the linear scaling of complexity with the number of transmit antennas $M_t$, which
makes NTCQ particularly attractive for massive MIMO systems for which conventional look-up based approaches are computationally
infeasible.


\subsection{Variations of NTCQ}\label{var_ntcq}
We can also construct several variations of NTCQ with minor tradeoffs between the total number of feedback bits, $B_{\mathrm{tot}}$, and performance.  We explain one of the variations briefly below.
\begin{itemize}
\item \textit{Variation: Fixing the starting state for the trellis search.}
\end{itemize}
Because NTCQ searches paths which start from every possible state in the first stage in the trellis, we need an additional $\log_2 S$ bits of feedback overhead to indicate the starting state of $\bp_{\mathrm{best}}$.  One variation is to fix the first state to eliminate these additional bits, so that the total feedback overhead incurred is exactly $BM_t$ bits.  We do incur a small performance loss by doing this, since allowing starting from different states effectively leads to considering more possible values of the scaling parameters $\alpha$ and $\theta$.  However, this loss becomes negligible as $M_t$ gets large (consistent with Theorem \ref{l_theorem}).

For other variations, we can fix the first entry of $\bc_{\mathrm{opt}}$ to a constant in the trellis search or adopt a tail-biting convolutional code.

\section{Advanced NTCQ Exploiting Channel Correlations}\label{advanced_NTCQ}
In practice, channels are temporally and/or spatially correlated.  In this section, we propose advanced NTCQ schemes that exploit these correlations to improve the performance or reduce the feedback overhead.

\subsection{Differential Scheme for Temporally Correlated Channels}\label{diff_sec}
A useful model of this correlation is the first-order Gauss-Markov process \cite{Tse:2006}
\begin{equation*}
\bh[k]=\eta
\bh[k-1]+\sqrt{1-\eta^2}\bg[k]
\end{equation*}
where $\bg[k]\in \mathbb{C}^{M_t}$ denotes the process noise, which is modeled as having i.i.d.\ entries distributed
with $\mathcal{CN}(0,1)$. We assume that the initial state
$\mathbf{h}[0]$ is independent of $\mathbf{g}[k]$ for all
$k$. The temporal correlation
coefficient $\eta$ $(0\leq \eta \leq 1)$ represents
the correlation between elements $h_{t}[k-1]$ and
$h_{t}[k]$ where $h_{t}[k]$ is the $t^{th}$ entry of
$\mathbf{h}[k]$.

If $\eta$ is close to one, two consecutive channels are highly correlated and the difference between the previous channel $\bh[k-1]$ and the current channel $\bh[k]$ might be small.  Differential codebooks in \cite{tm_correlated1,tm_correlated2,tm_correlated7,tm_correlated8,tm_correlated3,tm_correlated4,tm_correlated5,tm_correlated6} utilize this property to reduce the channel quantization error with an assumption that both the transmitter and the receiver know $\eta$ perfectly.  Most of the previous literature, however, focused on the case with a fixed and small number of transmit antennas and moderate feedback overhead, e.g., $M_t=4$ and $B_{\mathrm{tot}}=4$.  Therefore, we have to come up with a new differential feedback scheme to accommodate massive MIMO with large feedback overhead.

We denote $\bff[k-1]$ as the quantized beamforming vector at block $k-1$ and
\begin{equation*}
\bff_{\mathrm{opt}}[k]=\frac{\bh[k]}{\|\bh[k]\|_2}
\end{equation*}
as the unquantized optimal beamforming vector at time $k$.  In our differential NTCQ scheme, instead of quantizing $\bh[k]$ directly at time $k$, the receiver quantizes $\bff_{\mathrm{diff}}[k]$ which is given as
\begin{equation*}
\bff_{\mathrm{diff}}[k]=\left(\bI_{M_t}-\bff[k-1]\bff^H[k-1]\right)\bff_{\mathrm{opt}}[k].
\end{equation*}
Note that $\bff_{\mathrm{diff}}[k]$ is a projection of $\bff_{\mathrm{opt}}[k]$ to the null space of $\bff[k-1]$.  We let $\hat{\bff}_{\mathrm{diff}}[k]$ denote the quantized version of $\bff_{\mathrm{diff}}[k]$ by NTCQ with $\|\hat{\bff}_{\mathrm{diff}}[k]\|_2^2=1$. The receiver then constructs candidate beamforming vectors $\bff_{\bar{\alpha},\bar{\theta}}$ with weights $\bar{\alpha}\in\bar{\mathbb{A}}=\left\{\bar{\alpha}_1,\ldots,\bar{\alpha}_{K_{\bar{{\alpha}}}}\right\}$ and $\bar{\theta}\in\bar{\Theta}=\left\{\bar{\theta}_1,\ldots,\bar{\theta}_{K_{\bar{{\theta}}}}\right\}$ as
\begin{equation}\label{diff_combine}
\bff_{\bar{\alpha},\bar{\theta}}=\frac{\eta\bff[k-1]+\bar{\alpha}e^{j\bar{\theta}}\sqrt{1-\eta^2}\hat{\bff}_{\mathrm{diff}}[k]}
{\left|\left|\eta\bff[k-1]+\bar{\alpha}e^{j\bar{\theta}}\sqrt{1-\eta^2}\hat{\bff}_{\mathrm{diff}}[k]\right|\right|_2}.
\end{equation}
The receiver selects the optimal weights $\bar{\alpha}_{\mathrm{opt}}$ and $\bar{\theta}_{\mathrm{opt}}$ by optimizing
\begin{equation}\label{diff_beamforming}
\max_{\bar{\alpha}\in\bar{\mathbb{A}}}\max_{\bar{\theta}\in\bar{\Theta}}\left|\bar{\bh}^H[k]\bff_{\bar{\alpha},\bar{\theta}}\right|^2,
\end{equation}
and the final beamforming vector is given as 
\begin{equation*}
\bff[k]=\bff_{\bar{\alpha}_{\mathrm{opt}},\bar{\theta}_{\mathrm{opt}}}.
\end{equation*}
To construct candidate beamformaing vectors as in (\ref{diff_combine}), we have to define sets of weights $\bar{\mathbb{A}}$ and $\bar{\Theta}$.  It is easy to conclude that $\bar{\Theta}=[0,2\pi)$ because the quantization process uses beamformer phase invariance.  To derive the range of the set $\bar{\mathbb{A}}$, we make the following proposition.
\begin{proposition}
When $\eta\rightarrow 1$, the range of $\bar{\mathbb{A}}$ can be set as
\begin{equation}\label{a_bound}
\frac{1-\eta}{\sqrt{1-\eta^2}}\leq \bar{\alpha}\leq \frac{1+\eta}{\sqrt{1-\eta^2}}.
\end{equation}
\end{proposition}
\begin{IEEEproof}
First, we define $\bff_{\bar{\alpha},\bar{\theta}}^{\mathrm{nom}}$ as the numerator of \eqref{diff_combine} as
\begin{align*}
\bff_{\bar{\alpha},\bar{\theta}}^{\mathrm{nom}}&=\eta\bff[k-1]+\bar{\alpha}e^{j\bar{\theta}}\sqrt{1-\eta^2}\hat{\bff}_{\mathrm{diff}}[k].
\end{align*}
Then, the norm square of $\bff_{\bar{\alpha},\bar{\theta}}^{\mathrm{nom}}$ becomes
\begin{align*}
\|\bff_{\bar{\alpha},\bar{\theta}}^{\mathrm{nom}}\|_2^2&=\eta^2+\bar{\alpha}^2(1-\eta^2)\\
&\qquad+2\bar{\alpha}\sqrt{1-\eta^2}\mathrm{Re}\left\{e^{j\bar{\theta}}\bff^H[k-1]\hat{\bff}_{\mathrm{diff}}[k]\right\}.
\end{align*}
Because $-1\leq\mathrm{Re}\left\{e^{j\bar{\theta}}\bff^H[k-1]\hat{\bff}_{\mathrm{diff}}[k]\right\}\leq1$,
we have
\begin{equation}\label{diff_bound}
\left(\eta-\bar{\alpha}\sqrt{1-\eta^2}\right)^2\leq \|\bff_{\bar{\alpha},\bar{\theta}}^{\mathrm{nom}}\|_2^2\leq \left(\eta+\bar{\alpha}\sqrt{1-\eta^2}\right)^2.
\end{equation}

Note that $\bff^H[k-1]\hat{\bff}_{\mathrm{diff}}[k]\approx0$ with a good quantizer.  Moreover, with the assumption of a slowly varying channel which is typically assumed in the differential codebook literature, we approximate $\eta\approx1$. Then we have $\|\bff_{\bar{\alpha},\bar{\theta}}^{\mathrm{nom}}\|_2^2=1$, and plugging this into \eqref{diff_bound} gives the range of $\bar{\alpha}$ in \eqref{a_bound}.
\end{IEEEproof}
Note that the range in (\ref{a_bound}) can be further optimized numerically.  In Section \ref{diff_simulation}, we set $\frac{1-\eta}{\sqrt{1-\eta^2}}\leq \bar{\alpha}\leq \frac{1+\eta}{3\sqrt{1-\eta^2}}$ for simulation. Once the receiver selects the optimal weights $\bar{\alpha}_{\mathrm{opt}}$ and $\bar{\theta}_{\mathrm{opt}}$ by (\ref{diff_beamforming}), it feeds back $\hat{\bff}_{\mathrm{diff}}[k]$, $\bar{\alpha}_{\mathrm{opt}}$ and $\bar{\theta}_{\mathrm{opt}}$ to the transmitter over the feedback link and the transmitter reconstructs $\bff[k]$ as in (\ref{diff_combine}).  Additional feedback overhead caused by $\bar{\alpha}_{\mathrm{opt}}$ and $\bar{\theta}_{\mathrm{opt}}$ can be very small compared to the feedback overhead for $\hat{\bff}_{\mathrm{diff}}[k]$.  Simulation indicates that 1 bit for $\bar{\alpha}_{\mathrm{opt}}$ and 3 bits for $\bar{\theta}_{\mathrm{opt}}$ is sufficient to have near-optimal performance in a low mobility scenario.

\subsection{Adaptive Scheme for Spatially Correlated Channels}\label{adap_sec}
If the transmit antennas are closely spaced, which is likely for a massive MIMO scenario, channels tend to be spatially correlated and can be modeled as
\begin{equation*}
    \bh[k]=\bR^{\frac{1}{2}}\bh_w[k]
\end{equation*}
where $\bh_w[k]$ is an uncorrelated MISO channel vector with i.i.d. complex Gaussian entries and $\bR=E\left[\bh[k]\bh^H[k]\right]$ is a correlation matrix of the channel where expectation is taken over $k$.  We assume that $\bR$ is a full-rank matrix.  For spatially correlated MISO channels, codebook skewing methods were proposed in \cite{sp_correlated0,sp_correlated1,sp_correlated2} such that codewords in a VQ codebook are rotated and normalized with respect to $\bR$ to quantize only the local space of the dominant eigenvector of $\bR$.  It was shown in \cite{sp_correlated0,sp_correlated1,sp_correlated2} that this skewing method can significantly reduce the quantization error with the same feedback overhead.  With NTCQ, however, there are no fixed VQ codewords for channel quantization which precludes the normal approach for skewing.  Therefore, we propose the following method to mimic skewing with NTCQ for spatially correlated MISO channels.

We assume that both the transmitter and the receiver know $\bR$ in advance\footnote{In practice, the transmitter can acquire an approximate knowledge of $\bR$ by averaging $\bff[k]$, i.e., $\bR\approx E\left[\bff[k]\bff^H[k]\right]$ where expectation is taken over $k$.}. At the receiver side, $\bh_w[k]$ is obtained by decorrelating $\bh[k]$ with $\bR^{-\frac{1}{2}}$, i.e., 
\begin{equation*}
\bh_w[k]=\bR^{-\frac{1}{2}}\bh[k].
\end{equation*}
Then the receiver quantizes $\bh_w[k]$ with NTCQ and get $\hat{\bh}_w[k]$.  The receiver feeds back $\hat{\bh}_w[k]$, and the transmitter reconstructs $\bff[k]$ as
\begin{equation*}
\bff[k]=\frac{\bR^{\frac{1}{2}}\hat{\bh}_w[k]}{\left|\left|\bR^{\frac{1}{2}}\hat{\bh}_w[k]\right|\right|_2}.
\end{equation*}
This procedure effectively decouples the procedure of exploiting spatial correlation from that of quantization,
while providing the same performance gain as standard skewing of fixed codewords.

\section{Performance Evaluation}\label{simul}
In this section, we present Monte-Carlo simulation results to evaluate the performance of NTCQ in i.i.d. channels, temporally correlated channels, and spatially correlated channels.  In each scenario, we simulate the original NTCQ and its variation, differential NTCQ, and spatially adaptive NTCQ explained in Sections \ref{sec3}, \ref{diff_sec}, and \ref{adap_sec}, respectively.
We use the average beamforming gain in dB scale
\begin{equation*}
  J_{\mathrm{avg}}^{\mathrm{dB}}=10\log_{10}\left(E[|\bh^H\bff|^2]\right)
\end{equation*}
as a performance metric where the expectation is over $\bh$.

\subsection{i.i.d. Rayleigh fading Channels}\label{iid}
\begin{figure}[t]
        \centering\includegraphics[width=1\columnwidth]{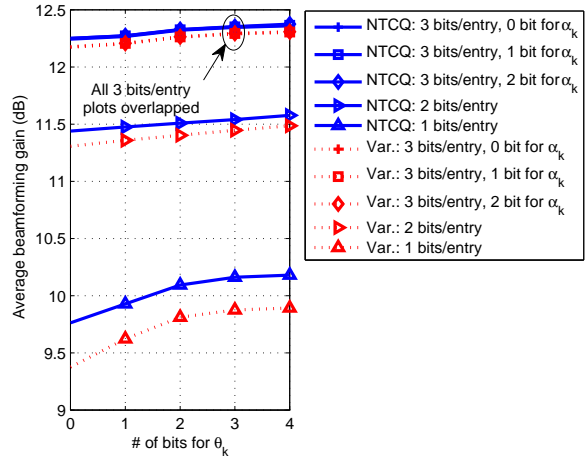}
        \caption{$J_{\mathrm{avg}}^{\mathrm{dB}}$ vs. different quantization levels of $\theta_k$ and $\alpha_k$ with $M_t=20$ in i.i.d. Rayleigh fading channels.}\label{iid_channel_diff_bit}
\end{figure}

For i.i.d. Rayleigh fading channels, $\bh[k]$ is drawn from i.i.d. complex Gaussian entries (i.e., $\bh[k]\sim\mathcal{CN}(\mathbf{0},\bI)$).  In Fig. \ref{iid_channel_diff_bit}, we first plot $J_{\mathrm{avg}}^{\mathrm{dB}}$ of NTCQ and its variation in i.i.d. channels with $M_t=20$ transmit antennas depending on different quantization levels for $\theta_k$ and $\alpha_k$.  Clearly, the variation of NTCQ gives strictly lower $J_{\mathrm{avg}}^{\mathrm{dB}}$ than the original NTCQ.  Note that it is enough to have $K_{\theta}=4$ (2 bits for $\theta_k$) for 1 bit/entry (QPSK) to achieve near-maximal performance of NTCQ and its variation.  Interestingly, we can fix $\alpha_k=1$ with 3 bits/entry (16QAM) for NTCQ and its variation without having any performance loss.  This is because when optimizing \eqref{16qam_optimizing}, it is likely to have $E\left[\|\bc_{\mathrm{opt}}\|_2^2\right]=1$ since the objective variable is the normalized channel vector $\bar{\bh}$ which has a unit norm, i.e., $\|\bar{\bh}\|_2^2=1$.  We fix $K_{\theta}=16$ (4 bits for $\theta_k$) for simulations afterward regardless of the number of bits per entry to have a fair comparison.  We also fix $\alpha_k=1$ for 3 bits/entry quantization.

\begin{figure}[t]
\centering\includegraphics[width=1\columnwidth]{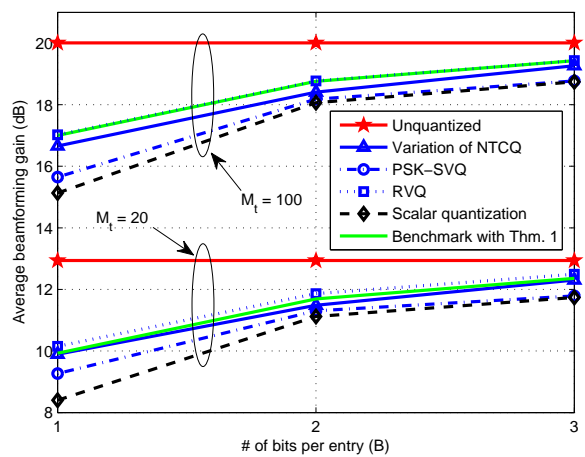}
\caption{$J_{\mathrm{avg}}^{\mathrm{dB}}$ vs. $B$ with $M_t=20$ and $100$  in i.i.d. Rayleigh fading channels.  PSK-SVQ is from \cite{QAM}.  All limited feedback schemes have the same $B_{\mathrm{tot}}$.}\label{iid_channel}
\end{figure}

In Fig. \ref{iid_channel}, we plot $J_{\mathrm{avg}}^{\mathrm{dB}}$ for variation of NTCQ (to have the same feedback overhead $B_{\mathrm{tot}}=BM_t$ with the other limited feedback schemes) as a function of the number of quantization bits per entry, $B$, in i.i.d. Rayleigh channel realizations.  We also plot $J_{\mathrm{avg}}^{\mathrm{dB}}$ for unquantized beamforming, RVQ, PSK-SVQ in \cite{QAM}, scalar quantization, and the benchmark from Theorem \ref{l_theorem} which is given as $M_t\left(1-2^{-B}\right)$ (in linear scale).  The performance of RVQ is plotted using the analytical approximation in \eqref{rvq} as $M_t\left(1-2^{-\frac{B_{\mathrm{tot}}}{M_t-1}}\right)$ (in linear scale), because it is computationally infeasible to simulate when the number of feedback bits grows large.  In scalar quantization, $B$ bits are used to quantize only the phase, not the amplitude, of each channel entry because the phase is generally more important than the amplitude in beamforming \cite{egt}.

As the number of feedback bits increases, the gap between the unquantized case and all limited feedback schemes decreases as expected.  RVQ gives the best performance among limited feedback schemes with the same number of feedback bits.  However, the difference between $J_{\mathrm{avg}}^{\mathrm{dB}}$ for RVQ and variation of NTCQ is small for all $B$.  The plots of the benchmark using Theorem \ref{l_theorem} well approximate $J_{\mathrm{avg}}^{\mathrm{dB}}$ of NTCQ for all $B$ and $M_t$, which shows the near-optimality of NTCQ.  Note that variation of NTCQ achieves better $J_{\mathrm{avg}}^{\mathrm{dB}}$ than PSK-SVQ regardless of $B$ and $M_t$, and the gap becomes larger as $M_t$ increases.  This gap comes from the coding gain of NTCQ.  As shown in Table \ref{bitperentry}, NTCQ can exploit $2^{B+1}$ constellation points while PSK-SVQ only utilizes $2^B$ constellation points with $B$ bits quantization per entry.  The coding gain of variation of NTCQ is around $0.25$ to $1$dB depending on $M_t$ and $B$.  Although we do not plot the performance of QAM-SVQ which relies on QAM constellations, it has the same structure as PSK-SVQ meaning that QAM-SVQ roughly experiences the same performance degradation compared to NTCQ.

\subsection{Temporally Correlated Channels}\label{diff_simulation}

To simulate the differential feedback schemes with the original NTCQ algorithm in temporally correlated channels, we adopt Jakes' model \cite{Prok} to generate the temporal correlation coefficient $\eta=J_{0}(2\pi f_D\tau)$, where $J_0(\cdot)$ is the 0th order Bessel function of the first kind, $f_D$ denotes the maximum Doppler frequency, and $\tau$ denotes the channel instantiation interval. We assume a carrier frequency of 2.5 $GHz$ and $\tau=5ms$.  We set the quantization level for the combiners $\bar{\theta}$ and $\bar{\alpha}$ in (\ref{diff_combine}) as 3 bits and 1 bit, respectively, which causes 4 bits of additional feedback overhead.

\begin{figure}[t]
\centering
\includegraphics[width=1\columnwidth]{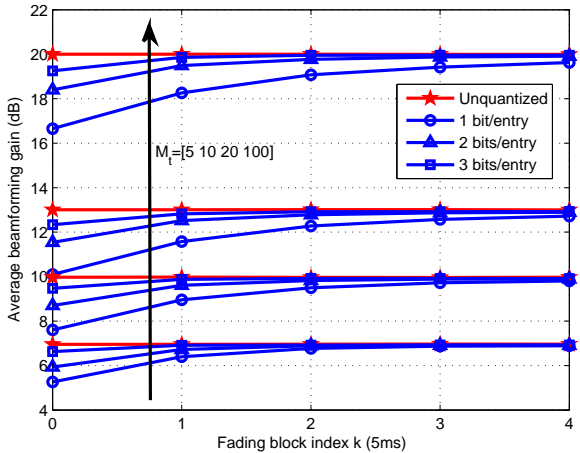}
\caption{$J_{\mathrm{avg}}^{\mathrm{dB}}$ vs. fading block index $k$ with $v=3km/h$ in temporally correlated channels. Without feedback delay.}\label{diff_09881_channel}
\end{figure}

In Fig. \ref{diff_09881_channel}, we plot the performance of the proposed differential NTCQ feedback schemes with the velocity $v=3km/h~(\eta=0.9881)$ assuming no feedback delay.  The differential NTCQ schemes, even with 1 bit/entry quantization, achieve almost the same performance as unquantized beamforming regardless of $M_t$.  Thus, if we can adjust the feedback overhead as a function of time, we can switch from NTCQ with 2 or 3 bits/entry quantization to 1bit/entry quantization in differential NTCQ to reduce the overall feedback overhead.

\begin{figure}
\centering
\includegraphics[width=1\columnwidth]{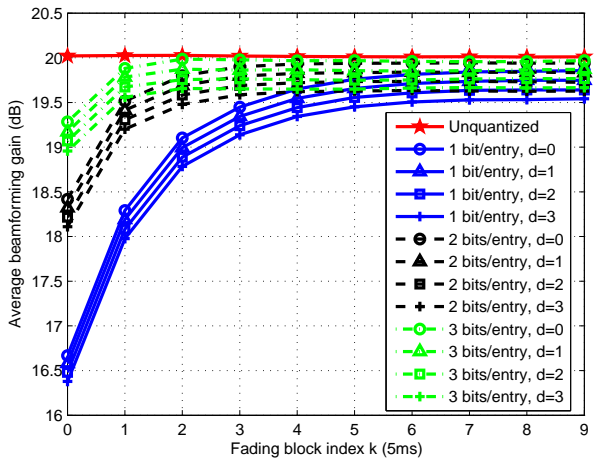}
\caption{$J_{\mathrm{avg-delay}}^{\mathrm{dB}}[d]$ vs. fading block index $k$ with $M_t=100$, $d$ blocks of feedback delay, and $v=3km/h$ in temporally correlated channels.}\label{diff_delay_channel}
\end{figure}

To see the effect of feedback delay in temporally correlated channels, we simulate the $M_t=100$ case with different numbers of delay $d$ measured in fading blocks (one fading block corresponds to $5ms$) in Fig. \ref{diff_delay_channel} such that
\begin{equation*}
J_{\mathrm{avg-delay}}^{\mathrm{dB}}[d]=10\log_{10}\left(E[|\bh^H[k]\bff[k-d]|^2)]\right).
\end{equation*}
It is shown that the effect of feedback delay is negligible, i.e., around $0.1$dB loss with one additional block delay for all cases, which confirms the practicality of the differential NTCQ scheme.  Moreover, we can reduce the frequency of the feedback updates to reduce the total amount of feedback overhead without significant performance degradation when the velocity of the receiver is low.


\subsection{Spatially Correlated Channels}\label{spatial_simulation}

To generate spatially correlated channels, we adopt the Kronecker model for the spatial correlation matrix $\bR$ which is given as $\bR=\bU\bSigma \bU^H$ where $\bU$ and $\bSigma$ are $M_t \times M_t$ eigenvector and diagonal eigenvalue matrices, respectively.  The performance of the adaptive scheme will highly depend on the amount of spatial correlation.  To see the effect of spatial correlation, we assume the eigenvalue matrix $\bSigma$ has a structure given by
\begin{equation*}
\bSigma=\diag\left\{\lambda_1,\frac{M_t-\lambda_1}{M_t-1}, \cdots, \frac{M_t-\lambda_1}{M_t-1}\right\}
\end{equation*}
where $1\leq\lambda_1<M_t$ is the dominant eigenvalue of $\bR$.  If $\lambda_1$ is small (large), the channels are loosely (highly) correlated in spatial domain.  Note that channels are i.i.d. when $\lambda_1=1$.
\begin{figure}[t]
\centering
\includegraphics[width=1\columnwidth]{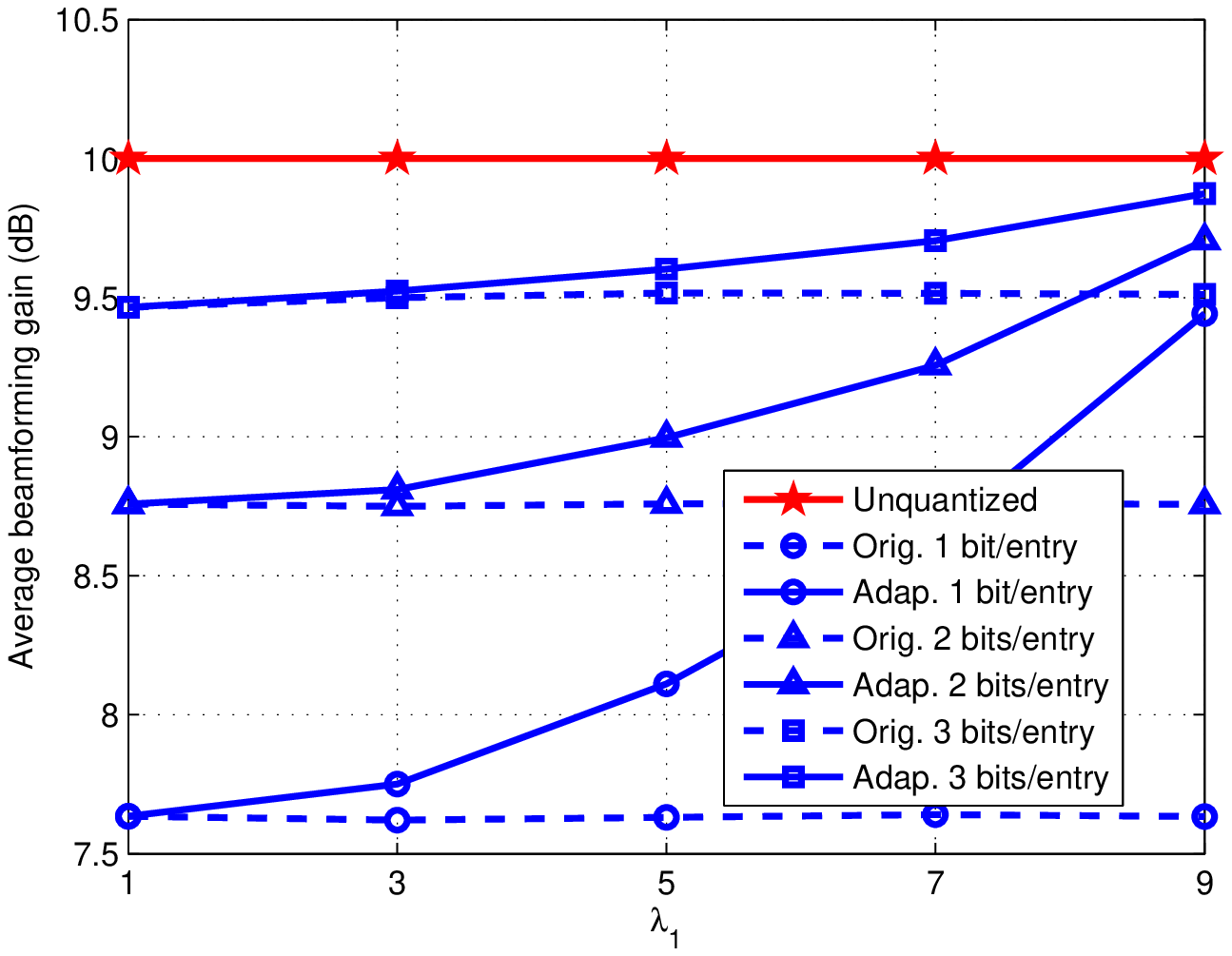}\\
\caption{$J_{\mathrm{avg}}^{\mathrm{dB}}$ vs. $\lambda_1$ with $M_t=10$ in spatially correlated channels.}\label{adap_10tx}
\end{figure}
\begin{figure}
\centering
\includegraphics[width=1\columnwidth]{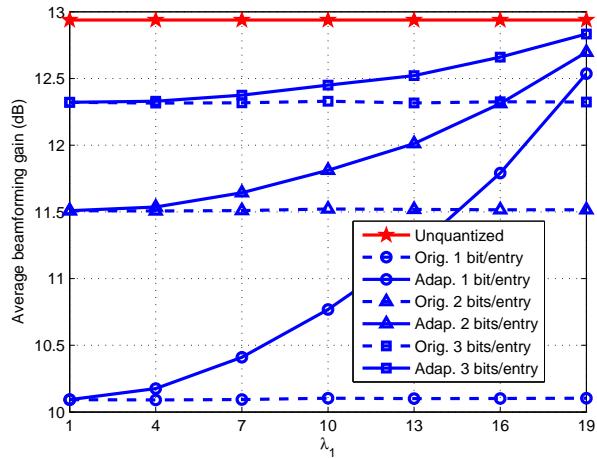}\\
\caption{$J_{\mathrm{avg}}^{\mathrm{dB}}$ vs. $\lambda_1$ with $M_t=20$ in spatially correlated channels.}\label{adap_20tx}
\end{figure}

In Fig. \ref{adap_10tx}, and \ref{adap_20tx}, we plot $J_{\mathrm{avg}}^{\mathrm{dB}}$ as a function of $\lambda_1$ for $M_t=10$ and $20$ cases.  The performance of spatially adaptive NTCQ become closer to that of unquantized beamforming as $\lambda_1$ increases with the same feedback overhead as original NTCQ.  This shows the effectiveness of the proposed adaptive NTCQ scheme for spatially correlated channels.

\section{Conclusions}\label{conclusion}

In this paper, we have proposed an efficient channel quantization method for massive MIMO systems employing limited feedback beamforming.
While the quantization criterion (maximization of beamforming gain or minimization of chordal distance) is associated with the Grassmann manifold, the key to the proposed NTCQ approach is to leverage efficient encoding (via the Viterbi algorithm) and codebook design
(via TCQ) in Euclidean space.  Efficient encoding relies on the mapping of quantization on the Grassmann manifold to noncoherent sequence detection and the near-optimal implementation of noncoherent detection using a bank of coherent detectors (i.e., Euclidean space quantizers).  Standard rate-distortion theory and asymptotic results for RVQ tell us that good Euclidean codebooks should work well in Grassmannian space. Our numerical
results show that the NTCQ provides better performance than uncoded schemes such as those considered in \cite{QAM}.

The advantages of NTCQ include flexibility and scalability in the number of channel coefficients: additional coefficients can be accommodated simply by increasing the blocklength, and the encoding complexity is linear in the number of transmit antennas.
It can also be easily modified to take advantage of channel conditions such as temporal and spatial correlations.
Our numerical results show that these advanced schemes can improve the performance significantly or reduce feedback overhead considerably depending on the system requirement.

While we have developed an efficient channel quantization method for massive MIMO systems, we note that limitations on feedback overhead would typically prevent scaling to an indefinitely large number of antennas.  However, the feedback overhead may be reasonable for the moderately large number of antennas (32 to 64) expected in initial deployments \cite{fdmimo}, and NCTQ represents a computationally efficient approach to generating such feedback.

Finally, in order to make FDD massive MIMO practical, it is also crucial to develop scalable sounding schemes for channel estimation.  Current sounding methods that transmit pilot signals from all transmit antennas using different time and/or frequency resources are not appropriate for massive MIMO systems because the pilot signals will dominate the downlink resources.  Initial work on this topic was conducted in \cite{cl_sounding} and extended in \cite{cl_sounding_journal}.

\bibliographystyle{IEEEtran}
\bibliography{ref}
\begin{biography}
{Junil Choi} (S'12) received his B.S. (with honors) and M.S. degrees from Seoul National University in Seoul, Korea in 2005 and 2007, respectively.  He is currently working toward the Ph.D. degree in the School of Electrical and Computer Engineering at Purdue University, West Lafayette, IN.  From 2007 to 2011, he was a member of the technical staff at Samsung Advanced Institute of Technology (SAIT) and Samsung Electronics in Korea, where he contributed advanced codebook and feedback framework designs to 3GPP LTE-Advanced and IEEE 802.16m standards.  His research interests are in the design and analysis of adaptive communication and massive MIMO systems.  Mr. Choi was a co-recipient of the 2008 Global Samsung Technical Conference Best Paper Award.  Recently, he was awarded the Michael and Katherine Birck Fellowship from Purdue University in 2011; and the Korean Government Scholarship Program for Study Overseas in 2011-2012.
\end{biography}

\begin{biography}
{Zachary Chance} (S'08-M'12) received the B.S. and Ph.D. degrees in electrical engineering from Purdue University, West Lafayette, IN, in 2007 and 2012, respectively.  In the summer and fall of 2011, he was with the Naval Research Laboratory, Washington, D.C., and MIT Lincoln Laboratory, Lexington, MA.  Since August 2012, he has been with MIT Lincoln Laboratory.  His research interests include wireless communications, feedback systems, and tracking.

Dr. Chance is an active reviewer for the IEEE Transactions on Wireless Communications, the IEEE Transactions on Communications, the IEEE Transactions on Signal Processing, and the EURASIP Journal on Wireless Communications.  He received the Ross Fellowship from Purdue University along with the Frederic R. Miller Scholarship, Mary Bryan Scholarship, Schlumberger Scholarship, and the Bostater-Tellkamp-Power Scholarship.

\end{biography}

\begin{biography}
{David J. Love} (S'98 - M'05 - SM'09) received the B.S. (with highest honors), M.S.E., and Ph.D. degrees in electrical engineering from the University of Texas at Austin in 2000, 2002, and 2004, respectively. During the summers of 2000 and 2002, he was with Texas Instruments, Dallas, TX. Since August 2004, he has been with the School of Electrical and Computer Engineering, Purdue University, West Lafayette, IN, where he is now a Professor and recognized as a University Faculty Scholar. He has served as an Associate Editor for both the IEEE Transactions on Communications and the IEEE Transactions on Signal Processing, and he has also served as a guest editor for special issues of the IEEE Journal on Selected Areas in Communications and the EURASIP Journal on Wireless Communications and Networking. His research interests are in the design and analysis of communication systems and MIMO array processing.  He has published over 120 technical papers in these areas and filed more than 20 U.S. patents, 17 of which have issued.

Dr. Love has been inducted into Tau Beta Pi and Eta Kappa Nu. Along with co-authors, he was awarded the 2009 IEEE Transactions on Vehicular Technology Jack Neubauer Memorial Award for the best systems paper published in the IEEE Transactions on Vehicular Technology in that year.  He was the recipient of the Fall 2010 Purdue HKN Outstanding Teacher Award and was an invited participant to the 2011 NAE Frontiers of Engineering Education Symposium.  In 2003, Dr. Love was awarded the IEEE Vehicular Technology Society Daniel Noble Fellowship.
\end{biography}

\begin{biography}
{Upamanyu Madhow} is Professor of Electrical and Computer Engineering at the University of California, Santa Barbara.  His research interests broadly span communications, signal processing and networking, with current emphasis on millimeter wave communication and bio-inspired approaches to networking and inference.  He received his bachelor's degree in electrical engineering from the Indian Institute of Technology, Kanpur, in 1985, and his Ph. D. degree in electrical engineering from the University of Illinois, Urbana-Champaign in 1990. He has worked as a research scientist at Bell Communications Research, Morristown, NJ, and as a faculty at the University of Illinois, Urbana-Champaign.  Dr. Madhow is a recipient of the 1996 NSF CAREER award, and co-recipient of the 2012 IEEE Marconi prize paper award in wireless communications. He has served as Associate Editor for the IEEE Transactions on Communications, the IEEE Transactions on Information Theory, and the IEEE Transactions on Information Forensics and Security. He is the author of the textbook Fundamentals of Digital Communication, published by Cambridge University Press in 2008.
\end{biography}
\end{document}